\documentclass[12pt]{article}
\begin{document}
\begin{titlepage}
\title{Kerr-Schild Tetrads and the Nijenhuis Tensor}

\author{J. W. Maluf$\,^{(a)(1)}$, F. L. Carneiro$\,^{(b)(2)}$, \\
\bigskip
S. C. Ulhoa$\,^{(c)(1)}$ and J. F. da Rocha-Neto$\,^{(d)(1)}$ \\
{(1)}\, Universidade de Bras\'{\i}lia, Instituto de F\'{\i}sica, \\
70.910-900 Bras\'{\i}lia DF, Brazil\\
(2)\, Universidade Federal do Norte do Tocantins, \\
77824-838 Aragua\'ina
TO, Brazil  }
\date{}
\maketitle
\begin{abstract}
We write the Kerr-Schild tetrads in terms of the flat space-time tetrads 
and of a (1,1) tensor $S^\lambda_\mu$. This tensor can be considered as a
projection operator, since it transforms (i) flat space-time tetrads into
non-flat tetrads, and vice-versa, and (ii) the Minkowski space-time 
metric tensor
into a non-flat metric tensor, and vice-versa. The $S^\lambda_\mu$ tensor 
and its inverse are constructed in terms of the standard null vector field
$l_\mu$ that defines the Kerr-Schild form of the metric tensor in general
relativity, and that yields black holes and non-linear gravitational waves 
as solutions of the vacuum Einstein's field
equations. We show that the condition for the vanishing of the Ricci tensor
obtained by Kerr and Schild, in empty space-time, is also a condition for
the vanishing of the Nijenhuis tensor constructed out of $S^\lambda_\mu$.
Thus, a theory based on the Nijenhuis tensor yields an important class of 
solutions of the Einstein's field equations, namely, black holes and 
non-linear gravitational waves. We also show that the present mathematical
framework can easily admit modifications of the Newtonian potential that may
explain the long range gravitational effects related to galaxy rotation 
curves. 
\end{abstract}
\thispagestyle{empty}
\vfill
\noindent (a) jwmaluf@gmail.com, wadih@unb.br\par
\noindent (b) fernandolessa45@gmail.com\par
\noindent (c) sc.ulhoa@gmail.com\par
\noindent (d) jfrocha@unb.br, jfrn74@yahoo.com \par
\end{titlepage}
\newpage

\section{The Kerr-Schild form of the metric tensor}
The Kerr-Schild form of the metric tensor \cite{KS} is a quite interesting 
construction that allows obtaining certain solutions of the vacuum 
Einstein's field equations. According to the book by Stephani {\it et. al.}
\cite{Stephani}, the Kerr-Schild ansatz was previously studied by Trautman 
\cite{Trautman}. It is given by

\begin{equation}
g_{\mu\nu}=\eta_{\mu\nu}+l_\mu l_\nu\,,
\label{1}
\end{equation}
where $\eta_{\mu\nu}$ is the metric for the Minkowski space-time in any 
coordinate system, and $l_\mu$ is a null vector field. 
In general, the Minkowski metric tensor is taken be 
$\eta_{\mu\nu}= (-1,+1,+1,+1)$ in a Cartesian coordinate system 
(which, however, is not necessarily rectangular). The form of the metric 
tensor (\ref{1}) allows one to easily transform covariant into
contravariant components, and vice-versa, by means of the flat Minkowski
metric tensor. The contravariant components of the metric tensor
are given by \cite{KS}

\begin{equation}
g^{\lambda\mu}=\eta^{\lambda\mu}-l^\lambda l^\mu\,.
\label{2}
\end{equation}
The vector $l_\mu$ is null with respect to both the metric tensors
$g_{\mu\nu}$ and $\eta_{\mu\nu}$ \cite{KS}, i.e., 

\begin{equation}
g^{\mu\nu}l_\mu l_\nu=\eta^{\mu\nu} l_\mu l_\nu=l^\mu l_\mu=0\,,
\label{3}
\end{equation}
where $l^\lambda=g^{\lambda \mu}l_\mu=\eta^{\lambda \mu}l_\mu$. In addition,
the determinant of the metric tensor satisfies \cite{KS} 
$(-g)=-\det(g_{\mu\nu})=1$. In the context of Riemannian geometry, 
the Kerr-Schild decomposition can be made only for type D and type O 
space-times.

Two gravitational field configurations that are of great relevance in general
relativity, black holes (Schwarzschild and Kerr)
and non-linear gravitational waves (pp-waves),
may be described by means of the Kerr-Schild ansatz 
\cite{Gursey,Ortaggio,Obukhov1,Obukhov2,Obukhov3}.
These are vacuum solutions of Einstein's field equations. 
The conditions under which matter fields may be included in 
the framework of the Kerr-Schild ansatz (as for instance, the electromagnetic
field) are discussed in Ref. \cite{Stephani} (see also Ref. \cite{Gurses}).
In particular, the metric tensor in the form of Eqs. (\ref{1}), (\ref{2})
may be treated as {\it exact linear perturbations} of the
Minkowski space-time \cite{Bini,Pujian}. Some recent investigations on 
the Kerr-Schild form are given in Refs. \cite{Aditya,Weijun}.

As noted by Kerr and Schild, the vacuum field equations 
$R_{\mu\nu}l^\mu l^\nu=0$ yield \cite{KS}

\begin{equation}
(l^\nu \partial_\nu l_\mu)(l^\lambda \partial_\lambda l^\mu)=0\,.
\label{4}
\end{equation}
However, from the derivative of the null condition $l^\mu l_\mu=0$ we obtain 
$l^\mu \partial_ \lambda l_\mu=0$, and consequently

\begin{equation}
l^\mu (l^\lambda \partial_ \lambda l_\mu)=0\,.
\label{5}
\end{equation}
Thus, the vector $l^\lambda \partial_\lambda l_\mu$ is null, from 
Eq. (\ref{4}), and is itself orthogonal to the null vector $l^\mu$.
Therefore we conclude that the vector $l^\lambda \partial_\lambda l_\mu$
besides being null, is collinear to the vector $l_\mu$, i.e.,

\begin{equation}
l^\lambda \partial_\lambda l_\mu = \sigma l_\mu\,.
\label{6}
\end{equation}
where $\sigma$ is a multiplicative factor. We are using mostly the notation
of Ref. \cite{KS}, with partial derivatives and adopting Cartesian 
coordinates, but the passage to arbitrary coordinates (and the
corresponding covariant derivatives in the flat space-time) is
straightforward \cite{Plebanski}. At this point, all indices may be raised
and lowered by means of the flat metric tensor $\eta_{\mu\nu}$.

The equations above express the basics of the Kerr-Schild ansatz. In the
following section we will construct the Kerr-Schild tetrads, and arrive at
the (1,1) tensor $S^\lambda_\mu$. Then in Section 3 we address the Nijenhuis
tensor, and conclude that the latter vanishes provided Eq. (\ref{6}) is
satisfied, i.e., as long as the Ricci tensor vanishes, according to Eq.
(\ref{4}). In Section 4 we show that a theory based on the Nijenhuis tensor
has freedom enough to admit solutions that describe deviations of the 
Newtonian potential that may explain the phenomenology related to galaxy
rotation curves. Presently, the investigation of this issue requires the
existence of some sort of dark matter, which so far has not been detected.
We conclude that the geometrical framework considered in this article is
quite rich and should be explored beyond the results presented here.

\section{Kerr-Schild tetrads}

The Kerr-Schild tetrads $e^a\,_\mu(x)$
that yield the metric tensors (\ref{1}) and (\ref{2})
are constructed by observing all issues of consistency. We first denote by
$E^a\,_\mu$ the flat space-time tetrads (i.e., tetrads for the flat Minkowski
space-time) in an arbitrary inertial state. The latin index $a$ is a SO(3,1) 
tangent space index. Given some matrix representation $\Lambda^a\,_b$ of the 
Lorentz Group, the tetrad fields $E^a\,_\mu$ transform as 
$\tilde{E}^a\,_\mu =\Lambda^a\,_b E^b\,_\mu$. We also denote the flat
tangent space-time metric tensor as $\eta_{ab}=(-1,+1,+1,+1)$. Therefore, 
the starting point for the construction of the 
Kerr-Schild tetrads is the expression

\begin{equation}
e^a\,_\mu= E^a\,_\mu + {1\over 2}l^a l_\mu\,,
\label{7}
\end{equation}
where 

\begin{equation}
l^a=E^a\,_\mu l^\mu\,.
\label{8}
\end{equation}
Tetrad fields of this type were recently considered in Ref. \cite{Frob}, in
the context of teleparallel gravity.
The flat space-time tetrad fields are required to satisfy

\begin{equation}
\eta_{ab}E^a\,_\mu E^b\,_\nu=\eta_{\mu\nu}\,.
\label{9}
\end{equation}
Thus, it follows from Eqs. (8) and (9) that

\begin{equation}
\eta_{ab}l^a l^b=\eta_{\mu\nu}l^\mu l^\nu\,.
\label{10}
\end{equation}
As a consequence of the expressions above, we have

\begin{equation}
\eta_{ab}e^a\,_\mu e^b\,_\nu= \eta_{\mu\nu}+l_\mu l_\nu =g_{\mu\nu}\,.
\label{11}
\end{equation}

The inverse tetrads are defined by

\begin{equation}
e_b\,^\lambda= E_b\,^\lambda -{1\over 2} l_b\, l^\lambda\,.
\label{12}
\end{equation}
where $l_b =E_b\,^\mu\,l_\mu$.
It is important to note that all indices of both $E^a\,_\mu$ and
$E_b\,^\lambda$ are raised and lowered by $\eta_{\mu\nu}$, $\eta_{ab}$ and
their inverses. The null vectors $l^\lambda$ and $l^a$ are related by
$l^\lambda=E_a\,^\lambda l^a$. The vector $l^a$ is also a null vector,
$l^a l_a=0$. This condition can be obtained directly from Eq. (\ref{10}).
The expressions so far obtained allow to verify the following orthogonality
properties, 

\begin{eqnarray}
E_b\,^\lambda E^b\,_\mu &=& \delta^\lambda_\mu \,,\label{13} \\ 
e_b\,^\lambda e^b\,_\mu &=& \delta ^\lambda_\mu\,. \label{14}
\end{eqnarray}
The relations below can be verified by simple, direct calculations and
ensure the consistency and validity of the whole formulation presented here:

\begin{eqnarray}
\eta^{ab}E_a\,^\mu E_b\,^\nu &=& \eta^{\mu\nu}\,, \label{15} \\
\eta^{ab}e_a\,^\mu e_b\,^\nu &=&\eta^{\mu\nu} -l^\mu l^\nu = 
g^{\mu\nu}\,, \label{16} \\
e^a\,_\mu e^b\,_\nu g^{\mu\nu}&=&\eta^{ab}\,. \label{17}
\end{eqnarray}

Now we return to Eqs. (\ref{7}) and (\ref{8}) and observe that after simple
rearrangements, the tetrad fields (\ref{7}) may be first rewritten as

\begin{equation}
e^a\,_\mu={1\over 2} E^a\,_\lambda(\delta^\lambda_\mu+
\eta^{\lambda \sigma}g_{\sigma\mu})\,.
\label{18}
\end{equation}
It is interesting to note, already at this point, (i) that the 
transformations properties of $e^a\,_\mu$ under Lorentz transformations are 
determined by the flat space-time tetrads $E^a\,_\mu$, and that (ii) the 
tetrad fields are written in terms of the metric tensors $\eta^{\lambda \mu}$
and $g_{\lambda\mu}$.
The inverse tetrads may be rewritten in a similar form,

\begin{equation}
e_a\,^\mu={1\over 2}E_a\,^\lambda(\delta^\mu_\lambda + 
\eta_{\lambda\rho}g^{\rho\mu})\,.
\label{19}
\end{equation}
By means of straightforward manipulations, we again rewrite the tetrad field 
$e^a\,_\mu$ given by Eq. (\ref{18}) in the form

\begin{equation}
e^a\,_\mu=E^a\,_\lambda S^\lambda_\mu\,,
\label{20}
\end{equation}
where the (1,1) tensor $S^\lambda_\mu$ is defined by

\begin{equation}
S^\lambda_\mu = \delta^\lambda_\mu + {1\over 2} l^\lambda l_\mu\,.
\label{21}
\end{equation}
It can be easily verified that the inverse of the tensor above is

\begin{equation}
(S^{-1})^\mu_\beta=\delta^\mu_\beta -{1\over 2} l^\mu l_\beta\,,
\label{22}
\end{equation}
i.e., $S^\lambda_\mu (S^{-1})^\mu_\beta =\delta ^\lambda_\beta$.
In terms of the inverse tensor $(S^{-1})^\mu_\beta$, we may rewrite the 
inverse tetrads (\ref{19}) in the form 

\begin{equation}
e_a\,^\mu=E_a\,^\lambda (S^{-1})_\lambda^\mu\,.
\label{23}
\end{equation}

The (1,1) tensors $S^\lambda_\mu$ and $(S^{-1})^\mu_\beta$ exhibit very 
interesting properties for both the metric tensor and for the tetrad fields.
First, for the metric tensor, it can be verified by simple calculations the
following relations,

\begin{eqnarray}
S^\mu_\rho\; S^\nu_\sigma \;\eta_{\mu\nu}&=& g_{\rho\sigma}\,, \label{24} \\
S^\mu_\rho \;S^\nu_\sigma \; g^{\rho\sigma}&=&\eta^{\mu\nu}\,, \label{25} \\
(S^{-1})^\mu_\rho \,(S^{-1})^\nu_\sigma \;\eta^{\rho\sigma}&=&
g^{\mu\nu}\,, \label{26} \\
(S^{-1})^\mu_\rho \,(S^{-1})^\nu_\sigma \;g_{\mu\nu} &=&
\eta_{\rho\sigma}\,. \label{27} 
\end{eqnarray}
As for the tetrad fields, besides equations (\ref{20}) and (\ref{23}) above,
we have

\begin{eqnarray}
E^a\,_\lambda &=& e^a\,_\mu\; (S^{-1})^\mu_\lambda\,, \label{28} \\
E_a\,^\rho&=& e_a\,^\mu\;S^\rho_\mu\,. \label{29}
\end{eqnarray}

In conclusion, we have the following rules, at least for the metric tensor
and for the tetrad fields. For a given tensor $S^\mu_\lambda$,
\begin{itemize}
\item the contravariant index $\mu$ converts flat space-time quantities
into non-flat quantities,
\item the covariant index $\lambda$ converts non-flat space-time quantities
into flat quantities.
\end{itemize}

As for the inverse tensor $(S^{-1})^\alpha_\beta$, the opposite takes place,

\begin{itemize}
\item the contravariant index $\alpha$ converts non-flat quantities
into flat quantities,
\item the covariant index $\beta$ converts flat space-time quantities
into non-flat quantities.
\end{itemize}

The general conclusion of the present analysis is that a (1,1) tensor of the
type $S^\mu_\lambda$ may play some role in gravity theories. On the other 
hand, such a (1,1) tensor plays a relevant role in the construction of the
Nijenhuis tensor. In the following section, we will investigate the 
implications of the $S^\mu_\lambda$ tensor given by Eq. (\ref{21}) in the
context of the Nijenhuis tensor.

\section{The Nijenhuis tensor}

The Nijenhuis tensor is a very interesting geometrical quantity since it is
entirely independent of any affine connection. It is defined by 
\cite{NJ,Nakahara}

\begin{equation}
N^\lambda_{\mu\nu}=S^\alpha_\mu \partial_\alpha S^\lambda_\nu-
S^\alpha_\nu \partial_\alpha S^\lambda_\mu -
S^\lambda_\alpha(\partial_\mu S^\alpha_\nu-\partial_\nu S^\alpha_\mu)\,.
\label{30}
\end{equation}
This tensor may be established in any differentiable manifold $M$ 
of arbitrary dimension D. To our knowledge, the Nijenhuis tensor has been 
considered in physics in two different
contexts. First, in the study of dynamical integral models 
\cite{Okubo1,Okubo2}. In this context, the vanishing of the Nijenhuis tensor
yields interesting 
properties in manifolds with dual symplectic structures. One of these 
properties is the emergence of a number of conserved quantities in
involution, that are necessary for the complete integrability of certain
dynamical systems \cite{Okubo1,Okubo2}. The Nijenhuis tensor 
is constructed out of a (1,1) tensor $S^\lambda_\mu$ not related to Eq. 
(\ref{21}), but to a dual symplectic structure of the manifold. 

In the context of complex manifolds, however, the vanishing of the Nijenhuis 
tensor is related to the existence of integrable almost complex structures 
\cite{Nakahara}. Some formulations of string theory demand that the 
internal six extra dimensions correspond to a complex manifold, endowed with 
a complex structure for which the Nijenhuis tensor vanishes \cite{Cardoso},
while the ordinary four dimensional space-time is identified with the Minkowski
space-time. Here, we present an alternative point of view and show that the 
Nijenhuis tensor may play some role in gravity.

Let us analyse the expression of the Nijenhuis tensor constructed out of 
the tensor $S^\lambda_\mu$ given by Eq. (\ref{21}). By just using the null
conditions $l^\mu\,l_\mu=0$ and $l^\mu\,\partial_\alpha l_\mu=0$, we obtain

\begin{equation}
N^\lambda_{\mu\nu}={1\over 2} \lbrack l^\lambda \,l_\mu 
(l^\alpha \partial_\alpha l_\nu)-
 l^\lambda \,l_\nu 
(l^\alpha \partial_\alpha l_\mu) \rbrack\,.
\label{31}
\end{equation}
Considering now the validity of Eq. (\ref{6}), which follows from the 
vanishing of the Ricci tensor in empty space-times, we see that 
the Nijenhuis tensor vanishes,

\begin{equation}
N^\lambda_{\mu\nu}=0\,.
\label{32}
\end{equation}
Therefore some solutions of Einstein's field equations in vacuum may be
obtained from a theory based on the Nijenhuis tensor, provided the tensor
$S^\lambda_\mu$ is given by Eq. (\ref{21}). The metric tensor is 
determined by Eq. (\ref{24}), in which case the geometrization 
of the space-time is achieved after solving some field equations in flat
space-time. Note that $R_{\mu\nu}l^\mu l^\nu=0$ does represent only one
field equation, i.e., it does not represent the entirety of the vanishing
of the Ricci tensor in vacuum, which is given by 10 field equations. By just
counting the number of field equations (again, in vacuum), it is clear that
the vanishing of the Nijenhuis tensor does not, in general, lead to the 
vanishing of the Ricci tensor. The point of contact between the vanishing of
both the Ricci and Nijenhuis tensor seems to be given by Eq. (\ref{6}) only.

It is not the purpose of the present article to establish a complete theory
for gravity entirely based on the Nijenhuis tensor, because of the 
limitation to vacuum solutions, or to solutions for which the energy-momentum
tensor for the matter fields satisfies $T_{\mu\nu}\,l^\mu l^\nu=0$ 
\cite{Stephani}. Of course, this limitation takes place in the context 
of Einstein's general relativity. The issue of the interaction of the tensor
field $S^\lambda _\mu$ with matter fields requires a quite deep analysis that
will not be carried out in this article. The possibility of a non-ordinary 
coupling of the Nijenhuis and/or the $S^\lambda_\mu$ tensors with the 
matter fields might indicate a further departure from the standard 
formulation of general relativity, that deserves a very careful 
investigation. In any case, we will easily address two 
theories constructed out of $N^\lambda_{\mu\nu}$ in flat space-time. In these
two theories, the action integral is varied with respect to the tensor field
$S^\lambda_\mu$, and both theories lead to the vanishing of the Nijenhuis 
tensor. The Minkowski metric tensor $\eta^{\mu\nu}$ is held fixed.
The first one is determined by the Lagrangian density

\begin{equation}
\label{33}
L_1=k\,\sqrt{-g}\,N^\mu\,N_\mu\,,
\end{equation}
where $k$ is a constant, $\sqrt{-g}$ refers to the flat space-time in 
arbitrary coordinates, $N_\mu=N^\lambda_{\lambda\mu}=S^\rho_\mu 
\partial_\rho S^\lambda_\lambda-S^\lambda_\rho \partial_\mu S^\rho_\lambda$ 
and  $N^\mu=\eta^{\mu\lambda}\,N_\lambda$. By neglecting the surface terms
that arise in the variation of $L_1$, we find that this
Lagrangian density yields the field equations

\begin{equation}
\label{34}
\sqrt{-g}\,S^\mu_\rho\,\partial_\lambda N^\lambda + 
\sqrt{-g}\,N^\mu\,\partial_\rho S^\lambda_\lambda
-\delta^\mu_\rho\,\partial_\lambda(\sqrt{-g}N^\nu S^\lambda_\nu)=0\,.
\end{equation}

The second theory is the Yang-Mills type theory, also 
in flat space-time, defined by the Lagrangian density

\begin{equation}
L_2=k\sqrt{-g}\,N^\alpha_{\beta\gamma}N^\lambda_{\mu\nu}\,
\eta_{\alpha\lambda} \eta^{\beta\mu}\eta^{\gamma\mu}\equiv 
k\sqrt{-g}\,N^{\mu\nu}_\lambda N^\lambda_{\mu\nu}\,,
\label{35}
\end{equation}
where we have defined $N^{\mu\nu}_\lambda = N^\alpha_{\beta\gamma}\,
\eta_{\alpha\lambda} \eta^{\beta\mu}\eta^{\gamma\mu}$;  $k$  and 
$\sqrt{-g}$ are the same as in $L_1$.
Again, we neglect the surface terms that arise from the 
variation of the Lagrangian density above. The field equations that follow
from $L_2$ are

\begin{eqnarray}
&{}&\sqrt{-g}N^{\mu\nu}_\lambda\,\partial_\alpha S^\lambda_\nu-
\sqrt{-g}N^{\lambda \nu}_\alpha \partial_\lambda S^\mu_\nu \nonumber \\
&{}&-
\partial_\lambda(\sqrt{-g}N^{\nu\mu}_\alpha\, S^\lambda_\nu)+
\partial_\lambda(\sqrt{-g}N^{\lambda\mu}_\nu\,S^\nu_\alpha)=0\,.
\label{36}
\end{eqnarray}
Obviously, $N^\lambda_{\mu\nu}=0$ is a solution of both Eqs. (\ref{34})
and (\ref{36}).

Unfortunately, it is not possible to establish a comparison of the
Lagrangian densities (\ref{33}) and (\ref{35}) with the Hilbert-Einstein
Lagrangian density. Likewise, it is not possible to relate Eqs. (\ref{34})
and (\ref{36}) to Einstein's equations in vacuum. Of course, there will 
always exist a relationship between Einstein's equation in terms of the null
vector $l_\mu$ with the field equations above, again in terms of the null 
vector $l_\mu$, but this relationship is too much restricted, and perhaps not
useful. A relationship between Einstein's equations with the field equations 
above, Eqs. (\ref{34}) and (\ref{36}), for the basic field variable 
$S^\lambda_\mu$, apparently does not exist.

One interesting feature of a theory based on the Nijenhuis tensor is that
the theory cannot be linearised, although, of course, the solutions may be
linearised in the sense of weak field limits, for instance. Let us consider
an expression for $S^\lambda_\mu$ in the form

\begin{equation}
\label{37}
S^\lambda_\mu = \delta^\lambda_\mu + \varepsilon \,h^\lambda_\mu \,
\end{equation}
where $\varepsilon$ is an infinitesimal parameter. It is easy to verify that

\begin{equation}
\label{38}
N^\lambda_{\mu\nu}(S)=\varepsilon^2 N^\lambda_{\mu\nu}(h)\,.
\end{equation}
Therefore, a theory based on the Nijenhuis tensor is definitely a non-linear
theory.

One straightforward consequence of expression (\ref{21}) for the tensor 
$S^\lambda_\mu$ is the following. The null vector $l^\mu$ may be
considered an eigenvalue of the tensor $S^\lambda_\mu$ in the sense that
$S^\lambda_\mu \,l^\mu=l^\lambda$. In fact, this relation may be generalised
to several products of tensor $S^\lambda_\mu$, as for instance
$S^\lambda_\sigma\,S^\sigma_\rho\,S^\rho_\mu\,l^\mu=l^\lambda$.
An arbitrary product of the tensor $S^\lambda_\mu$ yields

\begin{equation}
\label{39}
S^\lambda_{\alpha_1}\,S^{\alpha_1}_{\alpha_2}\, \cdots \,S^{\alpha_{n-1}}_\mu
=\delta^\lambda_\mu+{n\over 2}\, l^\lambda l_\mu\,.
\end{equation}

Before closing this section, we present an alternative expression for the
Nijenhuis tensor. We consider an arbitrary (1,1) tensor 
$S^\lambda_\mu$ and define the quantity ${\cal T}^\lambda_{\mu\nu}$ according 
to

\begin{equation}
\label{40}
{\cal T}^\lambda_{\mu\nu}=\partial_\mu\, {S^\lambda_\nu} -
\partial_\nu\, S^\lambda_\mu\,.
\end{equation}
This quantity was considered in Ref. \cite{Maluf}, in the analysis of a 
consistent theory for massless spin 2 fields. In terms of the expression 
above, it is possible to rewrite the Nijenhuis tensor as

\begin{equation}
\label{41}
N^\lambda_{\mu\nu}=-S^\lambda_\sigma\lbrack {\cal T}^\sigma_{\mu\nu}(S)+
S^\alpha_\mu\,S^\beta_\nu\,{\cal T}^\sigma_{\alpha\beta}(S^{-1})\rbrack\,.
\end{equation}
We see that ${\cal T}^\lambda_{\mu\nu}$ may be understood as a building 
block for the Nijenhuis tensor. Indeed, in Ref. \cite{Thompson} a very clear
exposition of complex manifolds is presented, which supports this 
interpretation. In summary, a complex structure
is determined by a real mixed tensor ${\bf J}$. In a $2n$ dimensional manifold,
this mixed tensor satisfies (in the notation of Ref. \cite{Thompson})
$J^P_N J^N_M=-\delta^P_M$ and, in this context, the Nijenhuis tensor is
defined by

\begin{equation}
\label{42}
N^P_{MN}=J^Q_M(\partial_Q J^P_N-\partial_N J^P_Q)
-J^Q_N(\partial_Q J^P_M-\partial_M J^P_Q)\,.
\end{equation}
The almost complex structure J defines a complex structure if and only if the
associated Nijenhuis tensor vanishes \cite{Thompson}.

If we restrict the tensor $S^\lambda_\mu$ to expression (\ref{21}), namely, 
$S^\lambda_\mu=\delta^\lambda_\mu + {1\over 2} l^\lambda l_\mu$, then it is
not difficult to show that 

\begin{equation}
\label{43}
N^\sigma_{\mu\nu}=-(\delta^\alpha_\mu \delta^\beta_\nu-
S^\alpha_\mu S^\beta_\nu)S^\sigma_\rho {\cal T}^\rho_{\alpha \beta}.
\end{equation}
Thus, we see that if ${\cal T}^\rho_{\alpha \beta}$ vanishes, then 
$N^\sigma_{\mu\nu}$ vanishes, but the opposite is not true. It seems to be
impossible to invert the expression above and write 
${\cal T}^\rho_{\alpha \beta}$ in terms of $N^\sigma_{\mu\nu}$.

\section{Modifications of the Newtonian Potential}
One interesting development of the mathematical framework determined by both
the Nijenhuis tensor and the Kerr-Schild tetrads is the freedom in modifying 
the Newtonian potential. We will show in this section that this freedom
accommodates modified long range Newtonian potentials that may explain certain
models of dark matter. Let us start by considering the line element for the
Kerr space-time given in both Refs. \cite{Bini,Plebanski}. More specifically,
we will consider Eq. (20.46) of Ref. \cite{Plebanski}. This is the form 
in which the Kerr metric first appeared in the literature. By making the 
angular momentum parameter $a=0$ in this equation, we obtain the line element
for the Schwarzschild space-time,

\begin{eqnarray}
ds^2&=&-dt^2+dx^2+dy^2+dz^2+{{2m}\over r}
(dt + {x\over r}dx +{y\over r}dy+{z\over r}dz)^2  \nonumber \\
&=& (\eta_{\alpha\beta}+l_\alpha l_\beta) dx^\alpha dx^\beta\,,
\label{44}
\end{eqnarray}
where $r^2=x^2+y^2+z^2$ and $m=GM/c^2$. 
Thus, the null vector $l_\alpha$ is identified as

\begin{equation}
\label{45}
l_\alpha=\biggl({{2m}\over r}\biggr)^{1/2}
(1, {x\over r}, {y\over r}, {z\over r})\,.
\end{equation}

One possible modification of the Newtonian potential was proposed in Refs.
\cite{Mashhoon1,Mashhoon2}, based on the analyses of galactic rotation curves
investigated in Ref. \cite{Kuhn}. According to 
Refs. \cite{Mashhoon1,Mashhoon2}, the standard Newtonian potential may be
modified by means of the prescription

\begin{equation}
\label{46}
\Phi_g=-{{GM}\over r} + 
{{GM}\over \lambda}\ln \biggl({r\over \lambda}\biggr),
\end{equation}
where the constant length $\lambda$ may be taken to be 
$\lambda \approx1\,kpc = 3260$ light years. Different values of $\lambda$ are
discussed in Ref. \cite{Kuhn}. Thus, the modification of the 
Newtonian potential is achieved by replacing

\begin{equation}
\label{47}
{{2m}\over r} \rightarrow {{2m}\over r} -
{{2m}\over \lambda} \ln\biggl( {r\over \lambda}\biggr)\,,
\end{equation}
in Eq. (\ref{45}). Of course, the value of $\lambda$ depends on the nature
and/or model of the galaxies in consideration, but as we will show below, 
the result to be presented here does not depend on the particular value of
$\lambda$. 

By defining a function $f(r)$ according to

\begin{equation}
\label{48}
f(r) = {{2m}\over r}- 
{{2m}\over \lambda} \ln \biggl( {r\over \lambda} \biggr)\,,
\end{equation}
we may define a modified form of the null vector $l_\alpha$ as

\begin{equation}
\label{49}
l_\alpha=\sqrt{f(r)}\biggl( 1, {x\over r}, {y\over r}, {z\over r}\biggr)\,.
\end{equation}

The crucial property of the null vector $l_\alpha$ that leads to the 
vanishing of the Nijenhuis tensor is Eq. (\ref{6}). Eq. (\ref{45}) above
satisfies this property. However, it is possible to show by simple 
calculations that Eq. (\ref{49}) also satisfies this property. Let us 
define the function $a(r)$ according to

\begin{equation}
\label{50}
a(r)= x\partial_x f+y\partial_y f + 
z\partial_z f \equiv x^i \partial_i f\,.
\end{equation}
It is easy to show that the equation 
$l^\lambda \partial_\lambda l_\mu = \sigma l_\mu$ holds true, where the 
multiplicative factor $\sigma(r)$ is given by

\begin{equation}
\label{51}
\sigma(r)= {{a(r)}\over {2r\sqrt{f}}}\,.
\end{equation}
Therefore, the equation 
$l^\lambda \partial_\lambda l_\mu = \sigma l_\mu$, which leads to the
vanishing of the Nijenhuis tensor, is verified irrespective of the 
expression of the function $f(r)$, which means that the standard Newtonian
potential is not fixed by a theory constructed out of the Nijenhuis tensor.
Obviously, the vanishing of the 10 components of the Ricci tensor does
not lead to the result above; rather, it leads the well known
textbook result given by the standard Newtonian potential characterized by
the constant mass parameter $m$.

In Section 3 of Ref. \cite{Mashhoon1} it is discussed the various forms that
the modified Newtonian potential may assume. It is also discussed, based on
previous studies (see references therein), that the Newtonian potential 
(better saying, the actual gravitational potential) should be deducted on 
empirical grounds. This is an ongoing debate, to which the present analysis
might contribute, since the function $f(r)$ is arbitrary, but restricted to
satisfy certain asymptotic boundary conditions fixed by the actual physical
configuration. However, it must be noted that the arbitrariness in the 
expression of the function $f(r)$ may be related to the fact that in nature
we do not have inertial reference frames. All physical frames in nature are
accelerated. Inertial reference frames are idealisations, which are
relevant of course, but these idealisations require a drastic simplification
of the space-time manifold.

The freedom in the determination of the Newtonian gravitational 
potential in the present formalism (which is expressed by the $g_{00}$ 
component of the metric tensor in general relativity) may be related to the
determination of the classical electromagnetic scalar potential 
$V(x)=A_0(x)$. In the Hamiltonian formulation of the classical 
electromagnetic field, in terms of the 4-vector potential $A_\mu$, one
finds that $A_0$ arises as a Lagrange multiplier for the Gauss law
$\nabla \cdot \vec{E}=0$, which is an elliptic differential equation and a
first class constraint of the theory. The
determination of the scalar potential $A_0$ is made only after the scalar 
charge density (source) $\rho(x)$ is established, after which $A_0$ is
given by the well known integral of $\rho(x)$. Likewise, it could be that 
the Newtonian potential in the present mathematical framework is fixed only
after the determination of both the matter distribution and the coupling of
the tensor field $S^\lambda_\mu$ with the matter fields, i.e., the coupling
of $S^\lambda_\mu$ with the matter energy momentum 4 vector 
$P^{\mu}_{matter}$ and/or with the matter energy-momentum tensor 
$T_{\mu\nu}$. As we mentioned before, this issue deserves further 
investigation.

\section{Comments}
In this article we have established tetrads for the Kerr-Schild form of the
metric tensor in general relativity.  The Kerr-Schild ansatz has been 
considered in the literature as a manifestation of the gravitational field on
a flat Minkowski background. In spite of the limitations of this formulation,
the Kerr-Schild ansatz and its consequences are very interesting. The tetrad 
fields obtained in the present analysis, Eqs. (\ref{20}) and (\ref{23}), are 
given by multiplication of the tetrad fields for the flat
space-time with a (1,1) tensor $S^\lambda_\mu$. The emergence of this tensor
immediately leads us to address the Nijenhuis tensor. This tensor
dispenses the consideration of any space-time affine connection,
and this feature simplifies the formulation of a gravitational theory.
The main conclusion of the present article is that the mathematical structure
determined by the Nijenhuis tensor and the associated Kerr-Schild tetrads 
constitutes a rich geometrical framework that deserves further investigation.

We have seen that the 
condition for the vanishing of the Ricci tensor in the context of general
relativity, as obtained by Kerr and Schild, and which leads to important 
gravitational field configurations, is also a condition for the vanishing of
the Nijenhuis tensor. Thus, the connection between gravitational field
configurations and the Nijenhuis tensor should be explored. In principle,
the latter tensor would lead to theories that yield
field equations in a flat space-time. By 
means of Eq. (\ref{24}), we cast the results obtained from $S^\lambda_\mu$ in
geometrical form. The analysis of field equations in flat space-time, and the
{\it a posteriori} geometrization by means of Eq. (\ref{24}), would greatly
simplify the analysis of gravitational field configurations. It must be noted
that in the context of a theory defined in flat space-time, the problem of
existence of essential singularities in general relativity may be put in a
different perspective. The singularities would be a consequence of the system
of differential equations in flat space-time, just like the 
singularities in Maxwell's theory, and not a feature of curved space-times,
or of spaces with curvature and/or torsion, i.e., they are not singularities 
of the space-time itself. Thus, the problem of existence of
singularities in the standard metrical formulation of general relativity
would be shifted from a conceptual problem to a mathematical, albeit 
important problem.

Finally, we concluded that the mathematical framework considered in this 
article may address the so called ``dark matter problem", in fact with no
dark matter at all. We have seen in Section 4 
that the vanishing of the Nijenhuis tensor
also admits modifications of the Newtonian potential that might explain 
the unresolved problem of the galaxies rotation curves. This is an open
problem, to which the present analysis could contribute.

\end{document}